\documentclass[twocolumn,preprintnumbers,superscriptaddress,amsmath,amssymb,aps,pre,10pt]{revtex4-1}

\usepackage{graphicx}
\usepackage{dcolumn}
\usepackage{bm}
\usepackage{color}
\usepackage{multirow}
\usepackage{listings}

\begin{document}


\title{Linear and nonlinear correlations in order aggressiveness of Chinese stocks}

\author{Peng Yue}
 \affiliation{School of Business, East China University of Science and Technology, Shanghai 200237, China} %
 \affiliation{Research Center for Econophysics, East China University of Science and Technology, Shanghai 200237, China} %

\author{Hai-Chuan Xu}
 \email{hcxu@ecust.edu.cn}
 \affiliation{School of Business, East China University of Science and Technology, Shanghai 200237, China} %
 \affiliation{Research Center for Econophysics, East China University of Science and Technology, Shanghai 200237, China} %

 \author{Wei Chen}
 \affiliation{Shenzhen Stock Exchange, 5045 Shennan East Road, Shenzhen 518010, China} %

 \author{Xiong Xiong}
 \affiliation{College of Management and Economics, Tianjin University, Tianjin 300072, China}
 \affiliation{China Center for Social Computing and Analytics, Tianjin University, Tianjin 300072, China}

 \author{Wei-Xing Zhou}
 \email{wxzhou@ecust.edu.cn}
 \affiliation{School of Business, East China University of Science and Technology, Shanghai 200237, China} %
 \affiliation{Research Center for Econophysics, East China University of Science and Technology, Shanghai 200237, China} %
 \affiliation{Department of Mathematics, East China University of Science and Technology, Shanghai 200237, China} %

\date{\today}

\begin{abstract}
  The diagonal effect of orders is well documented in different markets, which states that orders are more likely to be followed by orders of the same aggressiveness and implies the presence of short-term correlations in order flows. Based on the order flow data of 43 Chinese stocks, we investigate if there are long-range correlations in the time series of order aggressiveness. The detrending moving average analysis shows that there are crossovers in the scaling behaviors of overall fluctuations and order aggressiveness exhibits linear long-term correlations. We design an objective procedure to determine the two Hurst indexes delimited by the crossover scale. We find no correlations in the short term and strong correlations in the long term for all stocks except for an outlier stock. The long-term correlation is found to depend on several firm specific characteristics. We also find that there are nonlinear long-term correlations in the order aggressiveness when we perform the multifractal detrending moving average analysis.
\\\\
  {\textit{Keywords}}: Econophyscis; Detrending Moving Average Analysis; Fractal Analysis; Order aggressiveness; Stock Market.
\end{abstract}


\maketitle


\section{Introduction}
\label{S1:Introduction}

Stock markets are complex systems, in which traders exchange stock shares to move the price. In an order-driven market, traders submit different types of orders. Inpatient traders are liquidity takers, who submit orders in the opposite-side limit order book and consume orders. Patient traders are liquidity providers, who submit orders inside the spread or in the same-side limit order book. The orders of liquidity takers are more aggressive than that of liquidity providers. Hence, the aggressiveness of orders is an important quantity characterizing traders' willingness to trade in financial markets.

Empirical analysis in different markets unveiled the diagonal effect of order flows, which states that orders are more likely to be followed by orders of the same aggressiveness \cite{Biais-Hillion-Spatt-1995-JF,Majois-2010-EL}. The well-documented diagonal effect implies the presence of short-term correlations in order flows. A relevant literature reports the presence of long-term correlations (or long memory) in the sign time series of order flows \cite{Lillo-Farmer-2004-SNDE,Gu-Zhou-2009-EPL,Toth-Lemperiere-Deremble-deLataillade-Kockelkoren-Bouchaud-2011-PRX}. However, to our knowledge, it is still an open question if there are long-term correlations in the aggressiveness of orders. To answer this question, we perform fractal analysis and multifractal analysis to investigate the order flow data of 43 Chinese stocks. The fractal analysis shows that order aggressiveness exhibits linear long-term correlations. However, the multifractal analysis does not find convincing evidence of nonlinear long-range correlations in order aggressiveness.

Indeed, complex systems usually exhibit complex behavior characterized by long-term power-law correlations \cite{Sornette-2004}. A wealth of methods have been developed to investigate and determine the correlation strength in long-term correlated time series \cite{Taqqu-Teverovsky-Willinger-1995-Fractals,Montanari-Taqqu-Teverovsky-1999-MCM, Audit-Bacry-Muzy-Arneodo-2002-IEEEtit,Delignieres-Ramdani-Lemoine-Torre-Fortes-Ninot-2006-JMPsy,Kantelhardt-2009-ECSS}, including the Hurst analysis (or rescaled range analysis) \cite{Hurst-1951-TASCE,Mandelbrot-Wallis-1969b-WRR}, the fluctuation analysis (FA) \cite{Peng-Buldyrev-Goldberger-Havlin-Sciortino-Simons-Stanley-1992-Nature}, the detrended fluctuation analysis (DFA) \cite{Peng-Buldyrev-Havlin-Simons-Stanley-Goldberger-1994-PRE}, the discrete wavelet transform \cite{Kantelhardt-Roman-Greiner-1995-PA,Bunde-Bunde-Havlin-Roman-Goldreich-Schellnhuber-1998-PRL}, the wavelet transform module maxima (WTMM) approaches \cite{Holschneider-1988-JSP,Muzy-Bacry-Arneodo-1991-PRL,Bacry-Muzy-Arneodo-1993-JSP,Muzy-Bacry-Arneodo-1993-PRE,Muzy-Bacry-Arneodo-1994-IJBC, Audit-Bacry-Muzy-Arneodo-2002-IEEEtit}, the detrending moving average analysis (DMA) \cite{Alessio-Carbone-Castelli-Frappietro-2002-EPJB,Carbone-Castelli-2003-SPIE,Carbone-Castelli-Stanley-2004-PA,Carbone-Stanley-2004-PA,Carbone-Castelli-Stanley-2004-PRE,Xu-Ivanov-Hu-Chen-Carbone-Stanley-2005-PRE,Arianos-Carbone-2007-PA,Carbone-2009-IEEE}, to list a few. There are a lot of efforts trying to rank the performances of different methods \cite{Xu-Ivanov-Hu-Chen-Carbone-Stanley-2005-PRE,Oswiecimka-Kwapien-Drozdz-2006-PRE,Bashan-Bartsch-Kantelhardt-Havlin-2008-PA,Serinaldi-2010-PA, Jiang-Zhou-2011-PRE,Huang-Schmitt-Hermand-Gagne-Lu-Liu-2011-PRE,Bryce-Sprague-2012-SR,Grech-Mazur-2013-PA,Grech-Mazur-2013-PRE,Grech-Mazur-2015-APPA}. The conclusions are mixed and no clear-cut consensus has been reached, which is not unreasonable since different studies used different generators to synthesize time series with different lengths. Nevertheless, it is well accepted that DFA and DMA are ``The Methods of Choice'' in determining the Hurst index of time series \cite{Shao-Gu-Jiang-Zhou-Sornette-2012-SR}. Therefore, we adopt the DMA method in this work.

These methods have been generalized in many directions to study fractal objects in high dimensions \cite{Gu-Zhou-2006-PRE,Carbone-2007-PRE,AlvarezRamirez-Echeverria-Rodriguez-2008-PA,Turk-Carbone-Chiaia-2010-PRE}, multifractal time series \cite{Kantelhardt-Zschiegner-KoscielnyBunde-Havlin-Bunde-Stanley-2002-PA,Gu-Zhou-2010-PRE}, long-term power-law cross correlations between two time \cite{Jun-Oh-Kim-2006-PRE,Podobnik-Stanley-2008-PRL,Zhou-2008-PRE,Podobnik-Horvatic-Petersen-Stanley-2009-PNAS,Horvatic-Stanley-Podobnik-2011-EPL, Jiang-Zhou-2011-PRE,Kristoufek-2011-EPL,Wang-Shang-Ge-2012-Fractals,Oswiecimka-Drozdz-Forczek-Jadach-Kwapien-2014-PRE,Xie-Jiang-Gu-Xiong-Zhou-2015-NJP}, and long-term power-law partial cross correlations for multivariate time series \cite{Yuan-Fu-Zhang-Piao-Xoplaki-Luterbacher-2015-SR,Qian-Liu-Jiang-Podobnik-Zhou-Stanley-2015-PRE,Jiang-Zhou-Stanley-2017-Fractals,Jiang-Yang-Wang-Zhou-2017-FoP}. In this work, we adopt the multifractal detrending moving average analysis (MF-DMA) to research is there exists multifractal nature in the order aggressiveness time series.

The rest of this paper is organized as follows. Section \ref{S1:Data} describes the data under investigation. Section \ref{S1:DMA} reports the results of the detrending moving average analysis. We perform multifractal detrending moving average analysis in Section \ref{S1:MFDMA}. Section \ref{S1:Conclusion} concludes this paper.

\section{Data description}
\label{S1:Data}

The Chinese stock market is the largest emerging market in the world and became the second largest stock market after the USA market since 2009. It contains an A-share market and a much smaller B-share market. A company listed on the A-share market can also issue B-shares under certain conditions. We use the order flow data of 32 A-share stocks and 11 B-share stocks traded on the Shenzhen Stock Exchange (SZSE) of China in 2003. Our sample stocks belonged to the 40 constituent stocks of the SZSE component index in 2003. The SZSE is open for trading from Monday to Friday except the public holidays and other dates as announced by the China Securities Regulatory Commission. On each trading day, the market opens at 9:15 and entered the opening call auction till 9:30. The continuous double auction operates from 9:30 to 11:30 and 13:00 to 15:00. Only the order flows during the continuous double auction are considered in this work.

Orders can be classified into different types according to their aggressiveness \cite{Biais-Hillion-Spatt-1995-JF,Ranaldo-2004-JFinM}. We determine the order aggressiveness time series $a_i(t)$ for stock $i$ as follows. The absolute value of $a_i(t)$ is larger if the order $t$ is more aggressive and the sign of $a_i(t)$ indicates the direction of order $t$ such that $a_i(t)$ is positive for buy orders and negative for sell orders.

If $t$ is a partially filled buy order, $a_i(t)=5$, whose price is higher than the best ask price and whose quantity is greater than the amount of matched orders on the sell limit order book. If $t$ is a partially filled sell order, $a_i(t)=-5$, whose price is lower than the best bid price and whose quantity is greater than the amount of matched orders on the buy limit order book. These two types of orders are the most aggressive and have the highest immediate price impact \cite{Zhou-2012-QF,Zhou-2012-NJP,Xu-Jiang-Zhou-2017-IJMPB}.

If $t$ is a filled buy order, $a_i(t)=4$, whose price is not lower than the best ask price and whose quantity is no less than the amount of matched orders on the sell limit order book. If $t$ is a filled sell order, $a_i(t)=-4$, whose price is not higher than the best bid price and whose quantity is no less than the amount of matched orders on the buy limit order book. These two types of orders are also marketable orders and have a power-law immediate price impact \cite{Zhou-2012-QF,Zhou-2012-NJP,Xu-Jiang-Zhou-2017-IJMPB}.

If order $t$ is placed inside the spread such that its price is higher than the best bid price and higher than the best ask price, $a_i(t)=3$ for buy orders and $a_i(t)=-3$ for sell orders. If order $t$ is placed on the same best such that its price is equal to the best price on the same side, $a_i(t)=2$ for buy orders and $a_i(t)=-2$ for sell orders. If order $t$ is placed inside the same-side limit order book, $a_i(t)=1$ for buy orders and $a_i(t)=-1$ for sell orders.

\begin{figure}[htb]
  \centering
  \includegraphics[width=8.6cm]{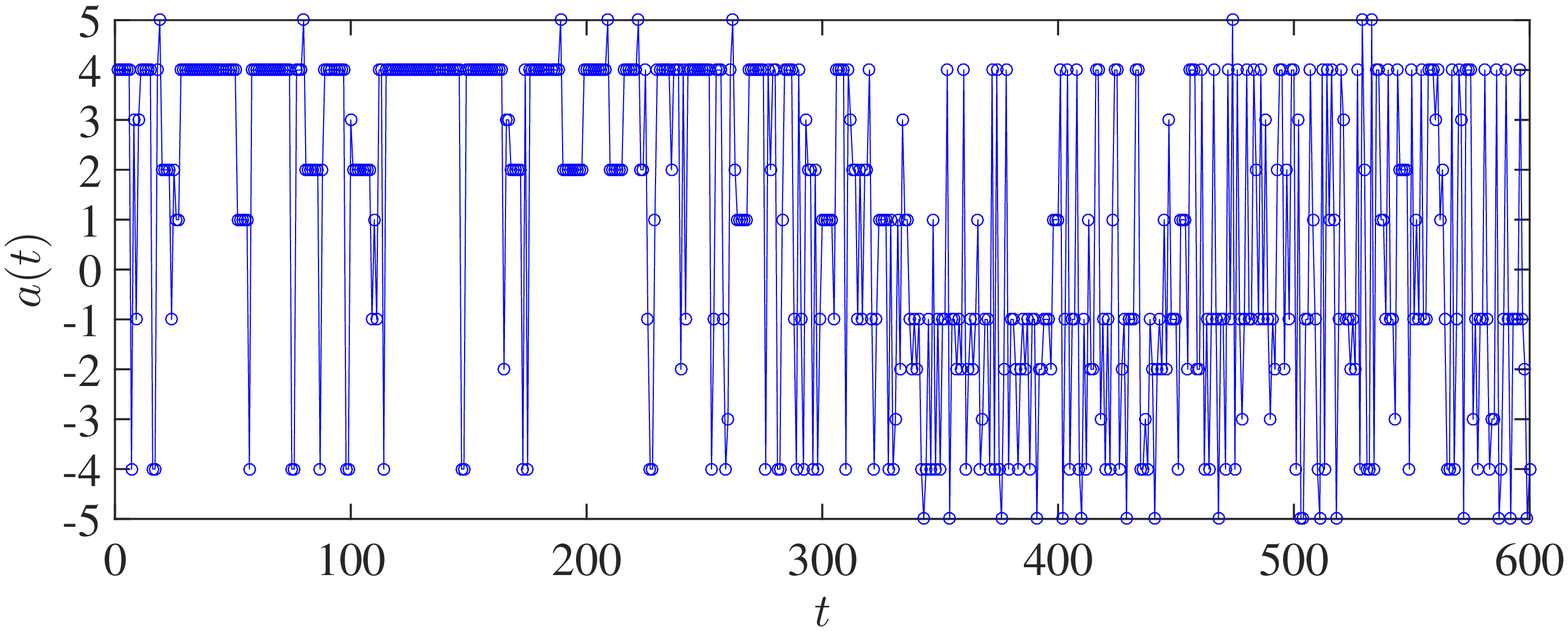}
  \includegraphics[width=8.6cm]{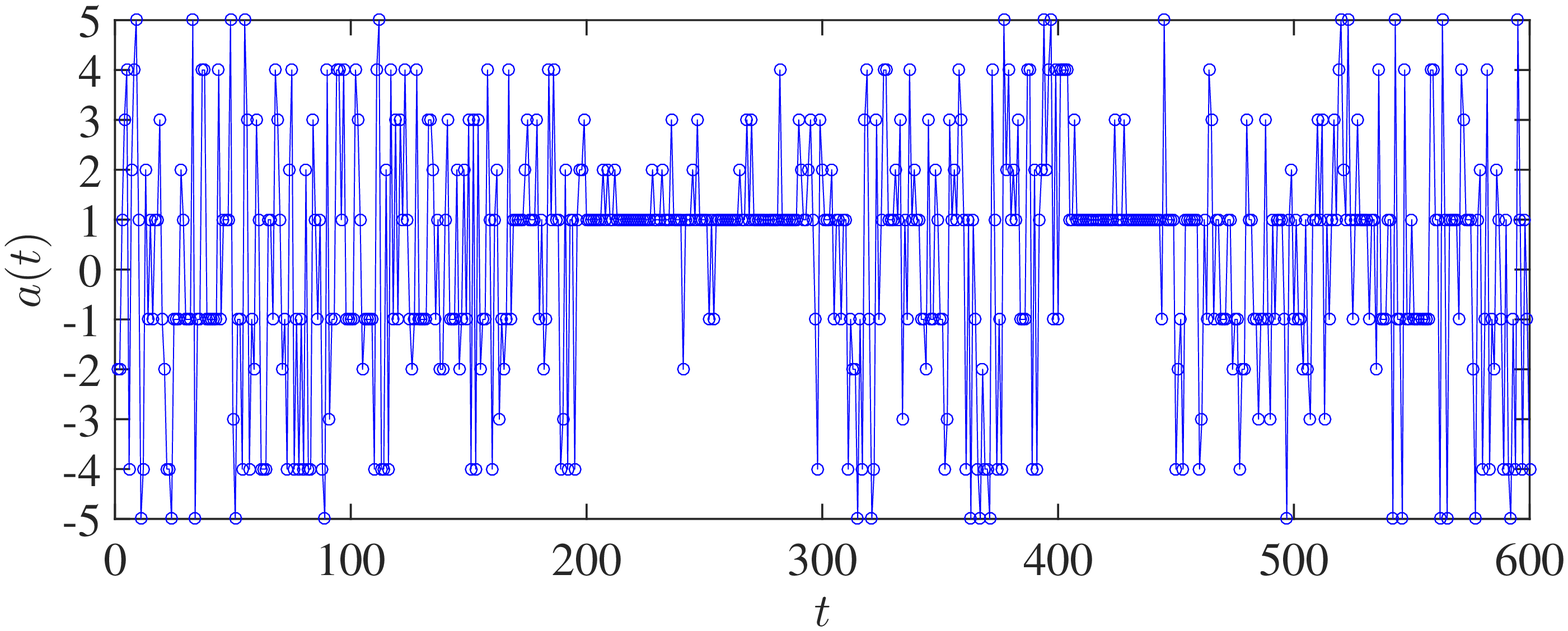}
  \caption{Example segments of the order aggressiveness time series $a_i(t)$ for an A-share stock 000720 (a) and a B-share stock 200541 (b).}
  \label{Fig:OrderType}
\end{figure}

Figure \ref{Fig:OrderType} shows two segments of the order aggressiveness time series $a_i(t)$ for an A-share stock (Code: 000720) and a B-share stock (Code: 200541). We choose these two segments to illustrate that the local patterns of $a_i(t)$ may change over time. The appearance of order bunches with the same aggressiveness is a signal of the diagonal effect. The diagonal effect may vary from stock to stock.

In total, we have 43 order aggressiveness time series $\{a_i(t): i=1,2,\cdots,43\}$ for 43 Chinese stocks. The average length of the time series is 802106, the maximum length is 3151313 (A-share stock 000001), and the minimum length is 67649 (B-share stock 200541). These time series are sufficiently long for DMA analysis and we do not need to consider the short series effect or the finite-size effect \cite{Zhou-2009-EPL,Zhou-2012-CSF,Shao-Gu-Jiang-Zhou-Sornette-2012-SR}.

\section{Linear long-term correlations}
\label{S1:DMA}

\subsection{The detrending moving average analysis}

We adopt the detrending moving average analysis to investigate the order aggressiveness for possible long-term correlations. The DMA method has been invented for this purpose and extensively studies \cite{Alessio-Carbone-Castelli-Frappietro-2002-EPJB,Carbone-Castelli-2003-SPIE,Carbone-Castelli-Stanley-2004-PA,Carbone-Stanley-2004-PA, Carbone-Castelli-Stanley-2004-PRE,Xu-Ivanov-Hu-Chen-Carbone-Stanley-2005-PRE,Arianos-Carbone-2007-PA,Carbone-2009-IEEE}. The procedure of the centred DMA method is briefly described below.

{\em{Step 1}}. Consider an order aggressiveness time series $a(t)$, $t=1,2,\ldots, N$, in which $t$ stands for the $t$-th order and the subscript $i$ for stock $i$ has been dropped. We construct the sequence of cumulative sums
\begin{equation}
 y(t)=\sum_{j=1}^{t}\left[a(j)-\langle{a}\rangle\right], ~~t=1, 2, \ldots, N,
 \label{Eq:DMA:y:t}
\end{equation}
where $N$ is the length of the time series $a(t)$ for a given stock and $\langle{a}\rangle$ is the sample mean of $a(t)$.

{\em{Step 2}}. Calculate the moving average function $\widetilde{y}(t)$ in a moving window of size $s$,
\begin{equation}
  \widetilde{y}(t)=\frac{1}{n}\sum_{k=-\lfloor(s-1)\theta\rfloor}^{\lceil(s-1)(1-\theta)\rceil}y(t-k),
  \label{Eq:DMA:y:t:tilde}
\end{equation}
where the operator $\lfloor{x}\rfloor$ is the largest integer not greater than $x$, $\lceil{x}\rceil$ is the smallest integer not smaller than $x$, and $\theta$ is the position parameter with the value varying in the range $[0,1]$. Here, $\theta=0$, $\theta=0.5$ and $\theta=1$ corresponding respectively to the backward, centred and forward DMA \cite{Gu-Zhou-2010-PRE}. We use $\theta=0.5$ in this work. Hence, Eq.~(\ref{Eq:DMA:y:t:tilde}) becomes
\begin{equation}
  \widetilde{y}(t)=\frac{1}{n}\sum_{k=-\lfloor(s-1)/2\rfloor}^{\lceil(s-1)/2\rceil}y(t-k),
  \label{Eq:DMA:y:t:tilde:centred}
\end{equation}

{\em{Step 3}}. Detrend the cumulative time series by removing the moving average function $\widetilde{y}(t)$ from $y(t)$, and obtain the residual time series $\epsilon(t)$ through the following equation
\begin{equation}
 \epsilon(t)=y(t)-\widetilde{y}(t),
 \label{Eq:DMA:epsilon}
\end{equation}
where $n-\lfloor(s-1)/2\rfloor\leqslant{i}\leqslant{N-\lfloor(s-1)/2\rfloor}$.

{\em{Step 4}}. Divide the residual time series $\epsilon(t)$ into $N_s$ disjoint segments with the same size $s$, where $N_s=\lfloor{N}/s-1\rfloor$. Each segment can be denoted by $\epsilon_v$ such that $\epsilon_v(t)=\epsilon(t+1)$ for $1\leqslant{t}\leqslant{s}$, where $l=(v-1)s$. The root-mean-square function $F_v(s)$ with the segment size $s$ can be calculated as follows
\begin{equation}
 F_v^2(s)=\frac{1}{n}\sum_{t=1}^{s}\epsilon_v^2(t).
 \label{Eq:DMA:Fv:s}
\end{equation}

{\em{Step 5}}. Determine the $q$th order overall fluctuation function $F_q(s)$ as follows,
\begin{equation}
  F_2(s) = \left\{\frac{1}{N_s}\sum_{v=1}^{N_s} {F_v^2(s)}\right\}^{\frac{1}{2}},
  \label{Eq:DMA:F2:s}
\end{equation}

{\em{Step 6}}. Determine the power-law relation between the function $F_q(s)$ and the size scale $s$ by varying the values of window size $s$, which reads
\begin{equation}
  F_2(s)\sim{s}^{H},
  \label{Eq:DMA:F2:s:H}
\end{equation}
where $H$ represents the DMA scaling exponent of the raw time series.

\subsection{Estimation of the Hurst indexes}

\begin{figure}[htb]
  \centering
\includegraphics[width=4cm]{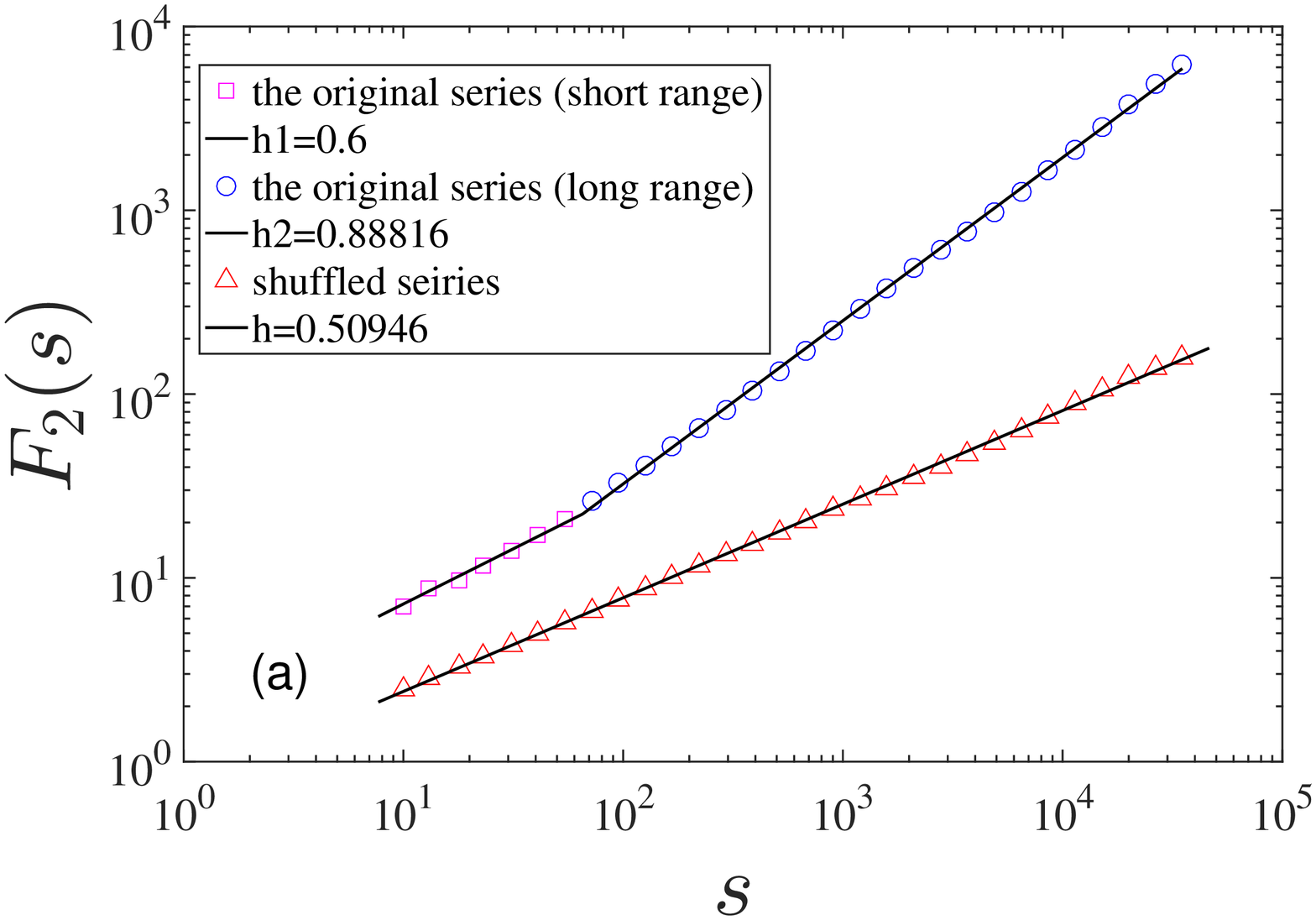}
\includegraphics[width=4cm]{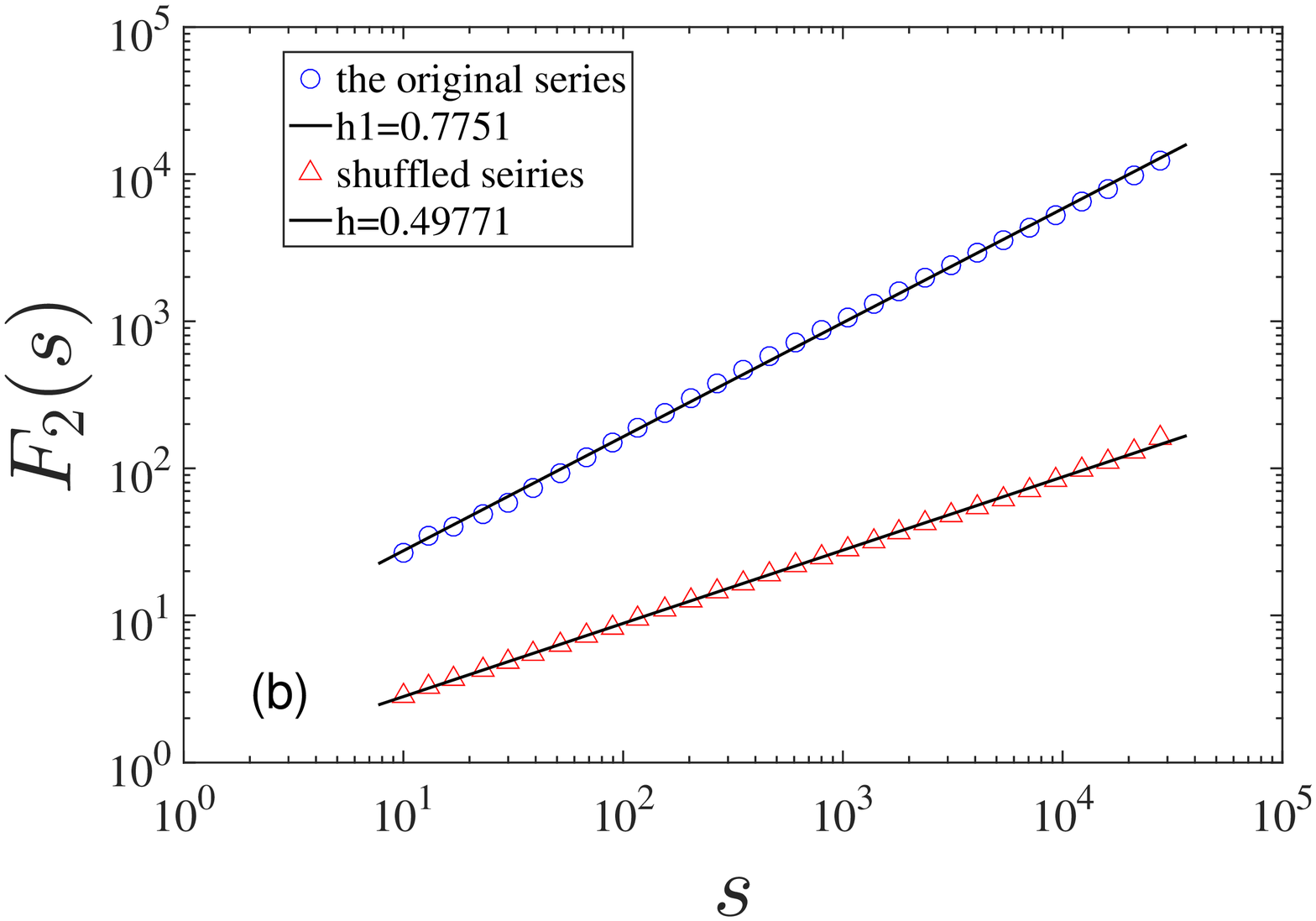}\vskip 0.2 cm
\includegraphics[width=4cm]{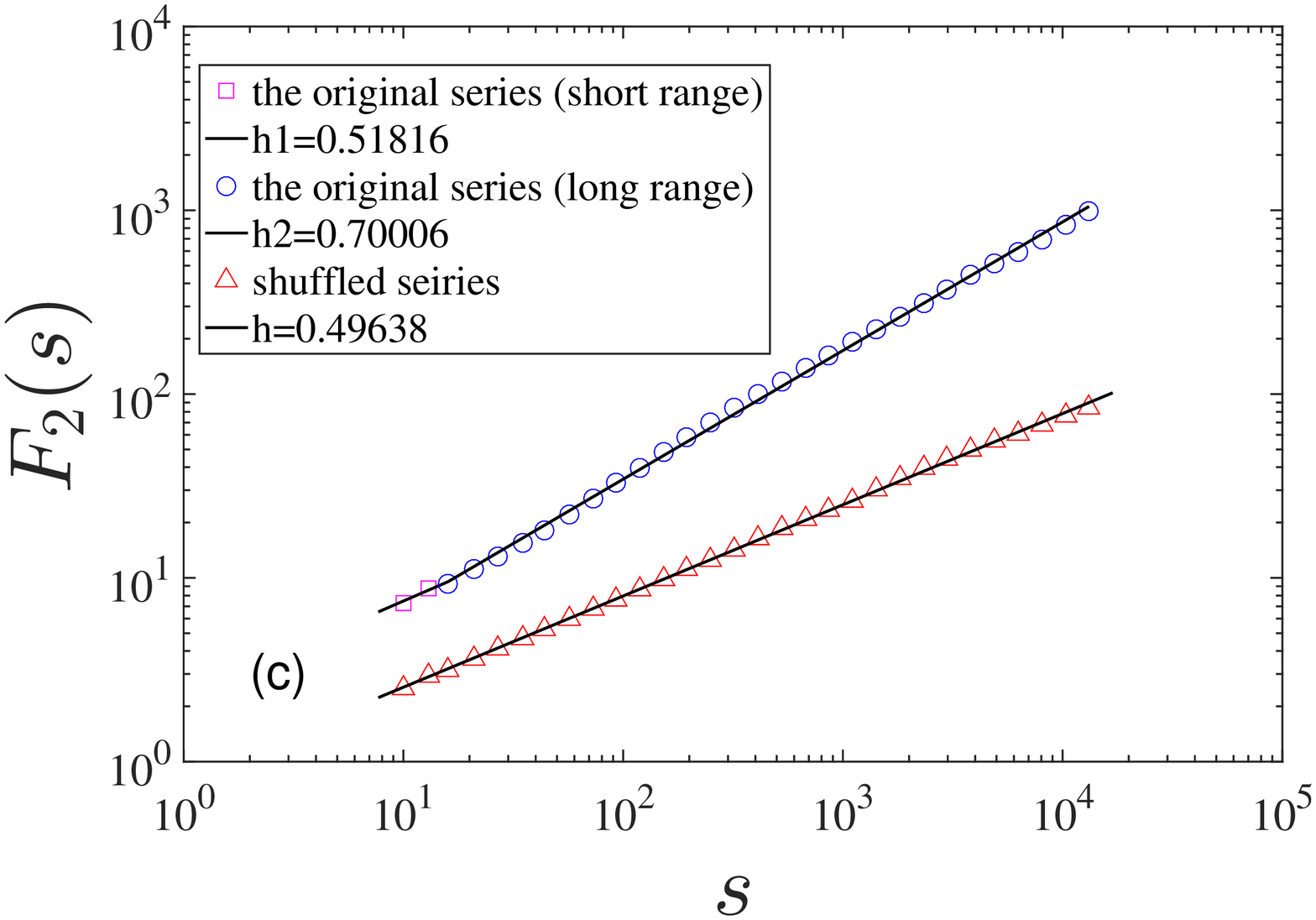}
\includegraphics[width=4cm]{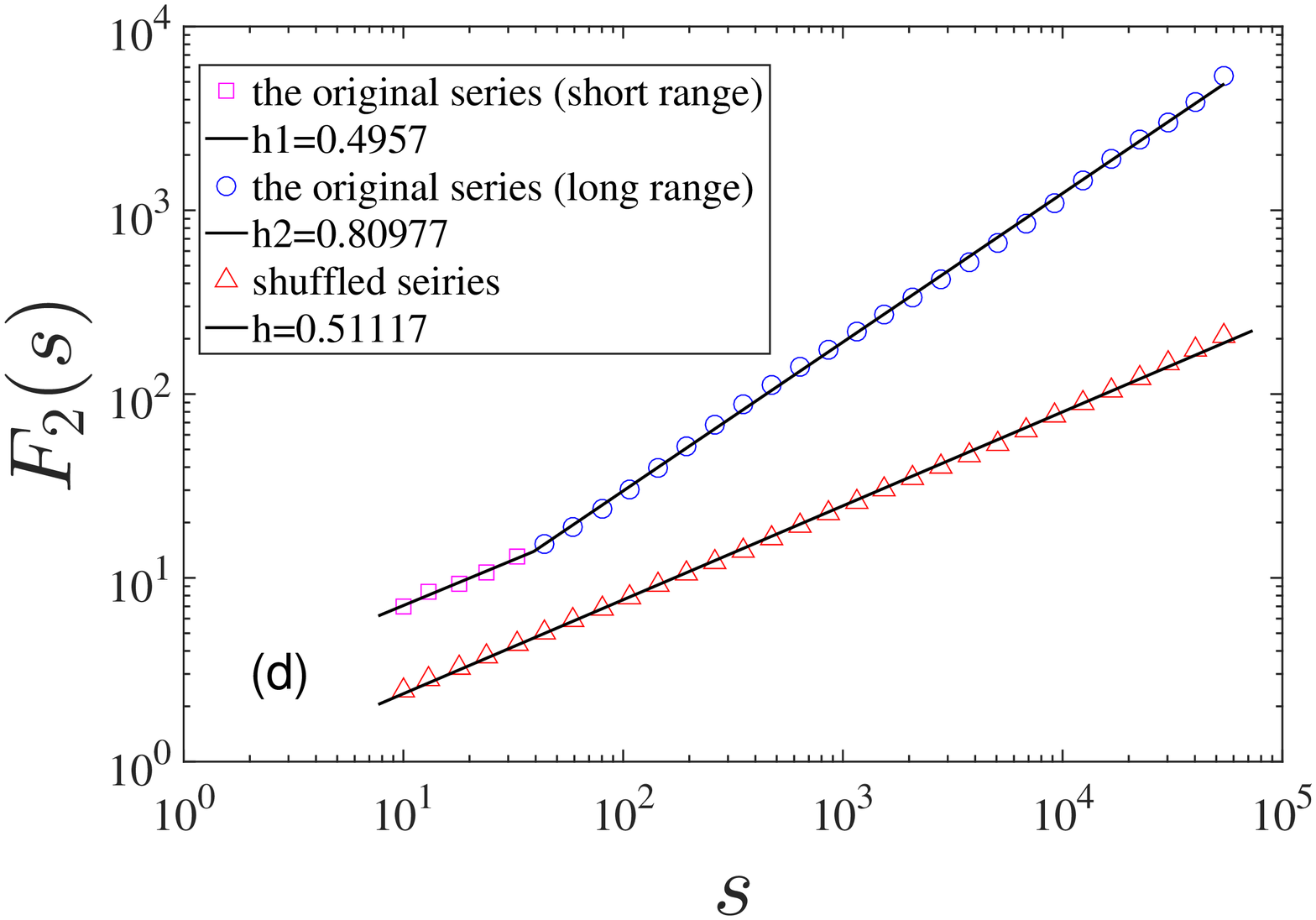}
  \caption{(color online) Power-law dependence of the overall fluctuation functions $F_2(s)$ with respect to the scale $s$. The straight lines are the best power-law fits to the data. The lower curves show the results for the shuffled time series, which have been translated vertically for better visibility. (a) A-share stock 000539. (b) A-share stock 000720. (c) B-share stock 200024. (d) B-share stock 200625.} 
\label{Fig:DMA:OrderType:FvS}
\end{figure}

Figure \ref{Fig:DMA:OrderType:FvS} illustrates the dependence of the overall fluctuation function $F_2(s)$ of the order aggressiveness time series as a function of the window size $s$ for two A-share stock and two B-share stocks. Roughly, we observer power-law relationships. However, there are also crossovers with $s_{\times}\in(10,100)$ such that
\begin{equation}
  F_2(s)\sim
  \begin{cases}
   {s}^{H_1}, ~~ s\leq{s}_{\times}\\
   {s}^{H_2}, ~~ s>{s}_{\times}
  \end{cases}
  \label{Eq:DMA:F2:sx:H1:H2}
\end{equation}
The DMA exponent is larger when the window size $s$ is larger than $s_{\times}$. The crossover behavior is ubiquitously reported in diverse systems and its possible origins contain additive noise and linear and non-linear trends \cite{Montanari-Taqqu-Teverovsky-1999-MCM,Kantelhardt-KoscielnyBunde-Rego-Havlin-Bunde-2001-PA,Hu-Ivanov-Chen-Carpena-Stanley-2001-PRE, Chen-Ivanov-Hu-Stanley-2002-PRE,Chen-Hu-Carpena-BernaolaGalvan-Stanley-Ivanov-2005-PRE,Ma-Bartsch-BernaolaGalvan-Yoneyama-Ivanov-2010-PRE, Arianos-Carbone-Turk-2011-PRE,Song-Shang-2011-Fractals,Zhao-Shang-Lin-Chen-2011-PA,Ludescher-Bogachev-Kantelhardt-Schumann-Bunde-2011-PA,Lin-Shang-2011-Fractals,Wang-Shang-Dong-2013-AMC,Shao-Gu-Jiang-Zhou-2015-Fractals}. For comparison, we also show the results for the shuffled time series. Basically, the DMA exponent is close to 0.5, ignoring the small curvature at large scale in Fig.~\ref{Fig:DMA:OrderType:FvS}(b).

We observe that all the 43 stocks exhibit the crossover behavior. In order to estimate the two DMA exponents $H_1$ and $H_2$, we need to determine first the crossover scale $s_{\times}$. We adopt an objective method for the simultaneous determination of $s_{\times}$, $H_1$ and $H_2$. We rewrite Eq.~(\ref{Eq:DMA:F2:sx:H1:H2}) as follows
\begin{equation}
  \ln{F_2(s)}=
   \begin{cases}
   c_1+H_1\ln{s}, & s\leq s_{\times}\\
   c_2+H_2\ln{s}, & s>s_{\times}
   \end{cases}
\end{equation}
Since the two straight lines should intersect at $s_{\times}$, that is, we have
\begin{equation}
  {c_1}+H_1\ln{s_{\times}} = {c_2}+H_2\ln{s_{\times}}.
  \label{Eq:DMA:Fit:constraint}
\end{equation}
To estimate the three parameters $s_{\times}$, $H_1$ and $H_2$, we minimize the following function
\begin{eqnarray}
  &O(s_{\times}, H_1, H_2, c_2)= \sum\limits_{s_j\leq{s_{\times}}}\left[\ln{F_2(s_j)}-H_1\ln{s_j}-{c_1}\right]^2\nonumber\\
   & + \sum\limits_{s_j\geq{s_{\times}}}[\ln{F_2(s_j)}-H_2\ln{s_j} -{c_2}]^2.
  \label{Eq:DMA:ObjFun}
\end{eqnarray}
where $c_1$ is not a free parameter since it is constrained by other four parameters according to Eq.~(\ref{Eq:DMA:Fit:constraint}).

\begin{table}[tb]
\centering
\caption{Estimated Hurst indexes $H_1$ and $H_2$ of the order aggressiveness time series in the short and long terms whose scaling ranges have a crossover at scale $s_{\times}$. The left panel and the right panel show respectively the results for A-share stocks and their corresponding B-share stocks (if available). Stock 000720 is an evident outlier without crossover. The $O_{\min}$ values are the minimized $O$ values.}
\smallskip
\scriptsize
\begin{tabular}{cccccccccccc}
\hline\hline
 \multicolumn{5}{c}{A-share stocks} & \multicolumn{5}{c}{B-share stocks}   \\\hline
  \cline{1-5} \cline{6-11} $code$ & $H_1$  & $H_2$ & $S_x$  & $O$ & $code$ & $H_1$ & $H_2$ & $S_x$ & $O$  \\\hline
000001 & 0.528 & 0.926 & 115.228 & 0.017 & & & & & \\
000002 & 0.515 & 0.909 & 72.144 & 0.024  &200002 & 0.522 & 0.754 & 30.148 &  0.017 &\\
000009 & 0.516 & 0.868 & 62.545 & 0.027 & & & & & \\
000012 & 0.513 & 0.812 & 39.783 & 0.040  &200012 & 0.541 & 0.713 & 22.000 &  0.024 &\\
000016 & 0.511 & 0.800 & 38.799 & 0.011  &200016 & 0.596 & 0.677 & 16.000 &  0.027 &\\
000021 & 0.513 & 0.825 & 46.482 & 0.026 & & & & & \\
000024 & 0.500 & 0.806 & 35.137 & 0.020  &200024 & 0.518 & 0.700 & 16.000 &  0.040 &\\
000027 & 0.534 & 0.894 & 63.898 & 0.009 & & & & & \\
000063 & 0.548 & 0.883 & 50.474 & 0.012 & & & & & \\
000066 & 0.514 & 0.826 & 42.541 & 0.020 & & & & & \\
000088 & 0.560 & 0.874 & 47.563 & 0.015 & & & & & \\
000089 & 0.528 & 0.857 & 44.324 & 0.014 & & & & & \\
000406 & 0.522 & 0.804 & 39.578 & 0.019 & & & & & \\
000429 & 0.481 & 0.747 & 24.955 & 0.019  &200429 & 0.534 & 0.741 & 30.150 &  0.011 &\\
000488 & 0.556 & 0.784 & 65.504 & 0.022  &200488 & 0.507 & 0.789 & 38.243 &  0.026 &\\
000539 & 0.600 & 0.888 & 65.402 & 0.007  &200539 & 0.528 & 0.805 & 46.609 &  0.020 &\\
000541 & 0.550 & 0.826 & 48.255 & 0.005  &200541 & 0.532 & 0.698 & 22.598 &  0.009 &\\
000550 & 0.518 & 0.819 & 37.722 & 0.024  &200550 & 0.590 & 0.746 & 21.981 &  0.012 &\\
000581 & 0.585 & 0.859 & 70.329 & 0.052  &200581 & 0.549 & 0.753 & 31.438 &  0.020 &\\
000625 & 0.531 & 0.828 & 53.185 & 0.018  &200625 & 0.496 & 0.810 & 39.217 &  0.015 &\\
000709 & 0.538 & 0.841 & 51.977 & 0.013 & & & & & \\
000720 & / & 0.775 &  / & / & & & & & \\
000778 & 0.522 & 0.832 & 53.877 & 0.019 & & & & & \\
000800 & 0.562 & 0.860 & 76.169 & 0.006 & & & & & \\
000825 & 0.531 & 0.862 & 62.382 & 0.039 & & & & & \\
000839 & 0.518 & 0.854 & 62.964 & 0.015 & & & & & \\
000858 & 0.519 & 0.838 & 61.020 & 0.014 & & & & & \\
000898 & 0.532 & 0.891 & 75.349 & 0.017 & & & & & \\
000917 & 0.476 & 0.775 & 26.300 & 0.014 & & & & & \\
000932 & 0.522 & 0.854 & 60.034 & 0.019 & & & & & \\
000956 & 0.520 & 0.819 & 50.880 & 0.020 & & & & & \\
000983 & 0.530 & 0.816 & 43.593 & 0.010 & & & & & \\
\hline\hline
\end{tabular}
\label{TB:H1:H2}
\end{table}

For each aggressiveness time series of the associated stock, we determine the three key parameters $s_{\times}$, $H_1$ and $H_2$, which are presented in Table~\ref{TB:H1:H2}. The left panel and the right panel show respectively the results for A-share stocks and their corresponding B-share stocks (if available). Note that only part of the A-share stocks have corresponding B-share stocks. All stocks exhibit a crossover in the scaling behavior except for stock 000720 (Shandong Luneng Taishan Cable Co. Ltd) which is an evident outlier. Indeed, stock 000720 was well recognized as a stock dominated by price manipulators and it exhibited distinct behaviors in many aspect: Its intertrade durations do not exhibit a crossover in the DFA scaling \cite{Jiang-Chen-Zhou-2009-PA}, its relative order prices have a different DFA scaling behavior during opening call auction \cite{Gu-Ren-Ni-Chen-Zhou-2010-PA}, its immediate price impact does not exhibit nice power-law scaling especially for filled buy trades \cite{Zhou-2012-QF}, its distribution of inter-cancellation durations behaves differently \cite{Gu-Xiong-Zhang-Zhang-Zhou-2014-FiP}, and so on.

Let's consider the short-term scaling behavior characterized by $H_1$. For A-share stocks, the minimum value is $H_{1,\min}=0.476$ for stock 000917, the maximum value is  $H_{1,\max}=0.600$ for stock 000539, and the average is $\langle{H_1}\rangle=0.529\pm0.026$. For B-share stocks, the minimum is $H_{1,\min} = 0.496$ for stock 200088, the maximum value is $H_{1,\max} = 0.596$ for stock 200009, and the average is $\langle{H_1}\rangle=0.538\pm0.031$. It shows that there is no significant correlations in the order aggressiveness time series in the short term and there is no evidence difference between A-shares and B-shares.

We then turn to the long-term scaling behavior characterized by $H_2$. For A-share stocks, the minimum value $H_{2,\min} = 0.747$ for stock 000429, the maximum value is $H_{2,\max} = 0.926$ for stock 000001, and the average is $\langle{H_2}\rangle=0.839\pm0.041$. For B-share stocks, the minimum value is $H_{2,\min} = 0.677$ for stock 200009, the maximum value is $H_{2,\max} = 0.810$ for stock 200088, and the average is $\langle{H_2}\rangle=0.744\pm0.044$. On average, the order aggressiveness of A-share stocks has stronger long-term correlations than B-share stocks. This can be attributed to the fact that the proportion of institutional traders is higher in the B-share market than the A-share market and retailer traders are more likely to herd.

\subsection{Determinants of the Hurst indexes}

We proceed to investigate possible firm-specific characters that might impact the variations of the Hurst indexes $H_1$ and $H_2$. The firm-specific characteristics we use include share trading volume $X_1$, dollar trading volume in RMB $X_2$, annual turnover rate based on full shares $X_3$, annual turnover rate based on tradable shares $X_4$, average daily turnover ratio based on full shares $X_5$, average daily turnover ratio based on tradable shares $X_6$, full shares $X_7$, tradable shares $X_8$, yearly return $X_9$, price earning ratio $X_{10}$, earnings per share $X_{11}$, return on equity $X_{12}$, operating profit per share $X_{13}$, net asset value per share $X_{14}$, and income per share $X_{15}$. A linear model is set up as follows
\begin{equation}
  H_{1,2} = X_0+ \sum_{i=1}^{15} \beta_iX_i + \epsilon,
  \label{Eq:H1:H2:Xi}
\end{equation}
We use the stepwise regression to determine which dependent arguments are statistically significant.

The results show that only the annual turnover rate based on full shares $X_3$ has a significant influence on $H_1$. The estimate of the coefficient is $\beta_{3} = -2.52\times10^{-4}$ with the $p$-value being 0.034. The adjusted $R^2$ is 0.083. It means that the Hurst index $H_1$ quantifying the short-term correlation decreases with increasing annual turnover rate. However, an increase of 10\% in the annual turnover rate will decrease $H_1$ by only $2.52\times10^{-5}$, which is negligible. We thus conclude that the investigated firm-specific characteristics do not influence the short-term correlations in the order aggressiveness time series. If we use only $X_6$ and $X_9$ to $X_{15}$ as independent arguments in the regression, no factors are statistically significant.

In contrast, the trading volume in RMB $X_2$, the average daily turnover ratio based on tradable shares $X_6$ and the income per share $X_{15}$ have significant impact on $H_2$ quantifying the long-term correlation. The estimates of the coefficients are  $\beta_{2}=0.061$, $\beta_{6}=-0.033$ and $\beta_{15}=-0.0012$ with the $p$-values being 0.000, 0.003 and 0.041 respectively. The adjusted $R^2$ is 0.776. Because the turnover is a better measure than share volume and dollar volume for trading activities \cite{Lo-Wang-2000-RFS} and the annual turnover rates are not accurate, we keep $X_6$ and remove $X_1$ to $X_5$, $X_7$ and $X_8$. We find that only $X_6$ and $X_{15}$ are statistically significant. The associated coefficients are $\beta_6=0.32$ and $\beta_{15}=-0.0032$ with the $p$-values being 0.020 and 0.0006. The adjusted $R^2$ is 0.320. The first conclusion is that the order aggressiveness has stronger long-term correlations if the turnover rate of the stock is higher reflecting that traders exhibit stronger imitations and herding. The result that $\beta_6=0.032$ is consistent with the results that $\beta_2=0.061$ and $\beta_6=-0.033$ such that $\beta_2+\beta_6=0.028$ in which all $X_i$ are included in the regression. The second conclusion is that the long-term correlation is stronger for stocks with lower income per share, which reflects the irrational trading behavior of Chinese traders who like to speculate low-performance stocks.



\section{Nonlinear long-range correlations}
\label{S1:MFDMA}

In order to check if there are any nonlinear long-range correlations in the order aggressiveness time series, we perform the multifractal detrending moving average (MFDMA) analysis \cite{Gu-Zhou-2010-PRE}. The MFDMA analysis is a extension of the DMA approach by generalizing the overall fluctuation function $F_2(s)$ to $F_q(s)$ of different orders, described as follows.

{\em{Step 5}}. Determine the $q$th order overall fluctuation function $F_q(s)$ as follows,
\begin{equation}
  F_q(s) = \left\{\frac{1}{N_s}\sum_{v=1}^{N_s}\left[F_v(s)\right]^q\right\}^{\frac{1}{q}},
  \label{Eq:1ddma:Fqs}
\end{equation}
where $q$ can take any real value except for $q=0$. When $q=0$, an application of L'H\^{o}spital's rule results in
\begin{equation}
  F_0(s) = \exp\left\{\frac{1}{N_s}\sum_{v=1}^{N_s}{\ln[F_v(s)]}\right\}.
  \label{Eq:1ddma:Fq0}
\end{equation}

{\em{Step 6}}. Determine the power-law relation between the function $F_q(s)$ and the size scale $s$, which reads
\begin{equation}
  F_q(s)\sim{s}^{h(q)}.
  \label{Eq:DMA:m:h}
\end{equation}

\begin{figure}[htb]
  \centering
  \includegraphics[width=4cm]{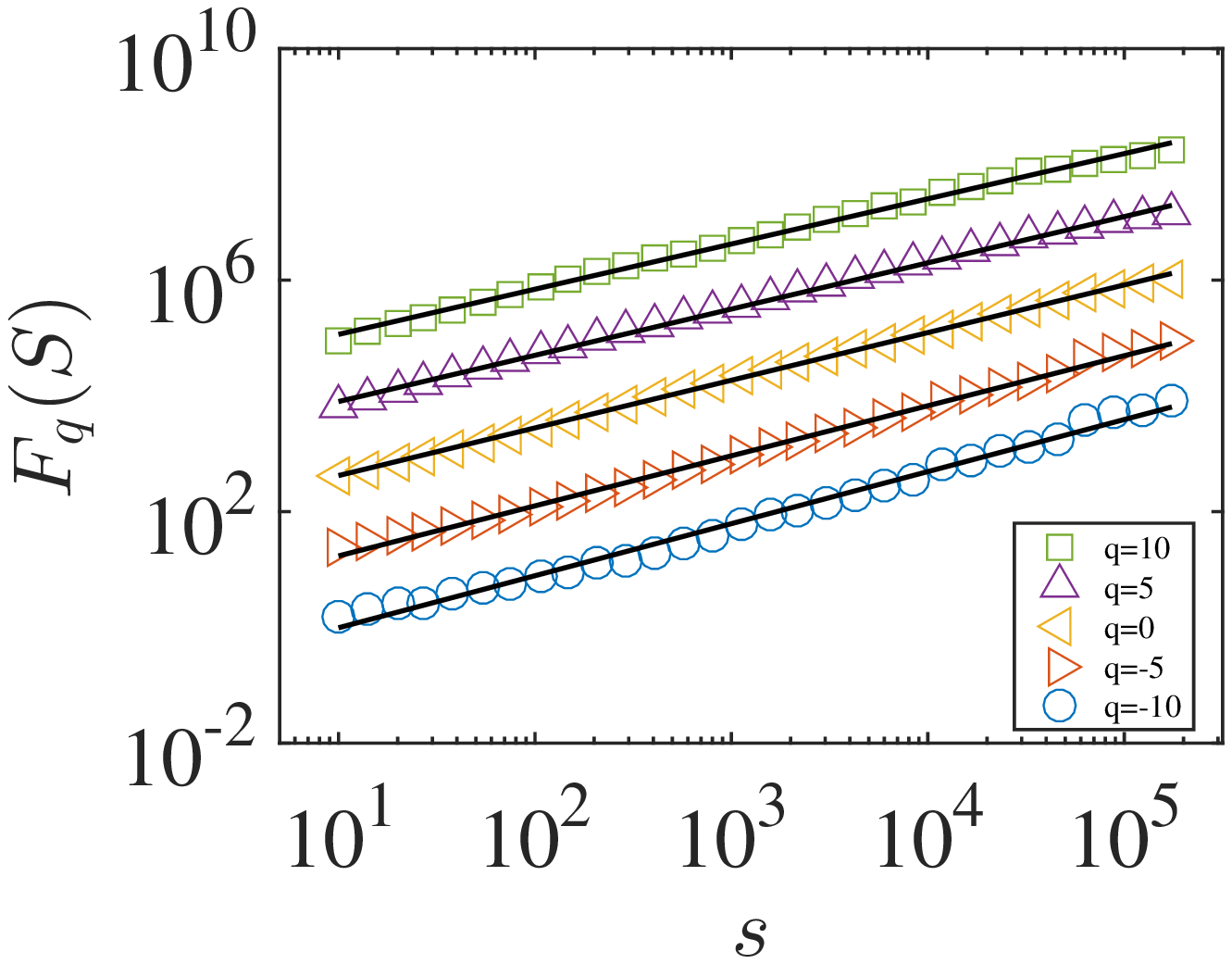}
  \includegraphics[width=4cm]{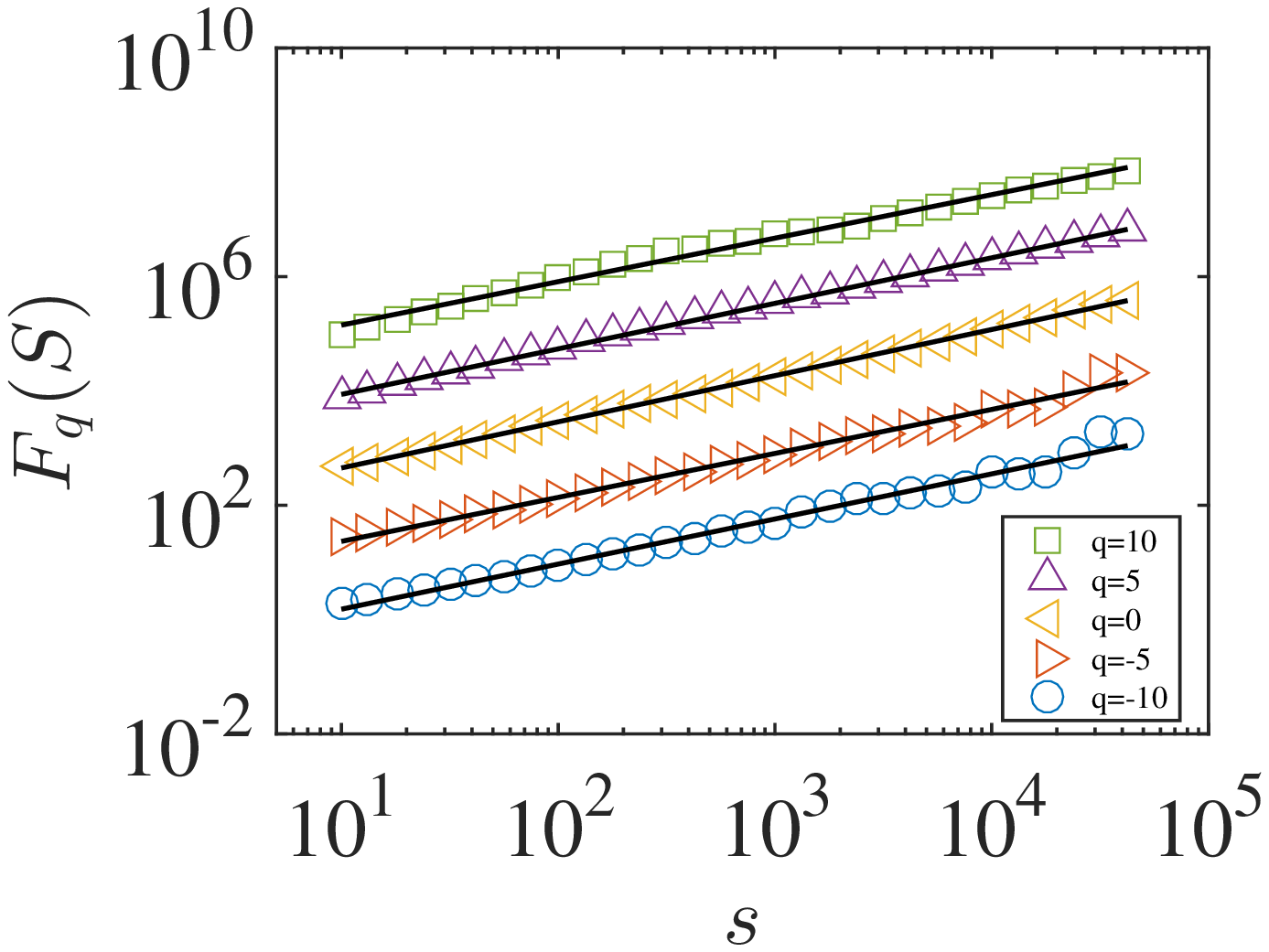}\vskip 0.3 cm
  \includegraphics[width=4cm]{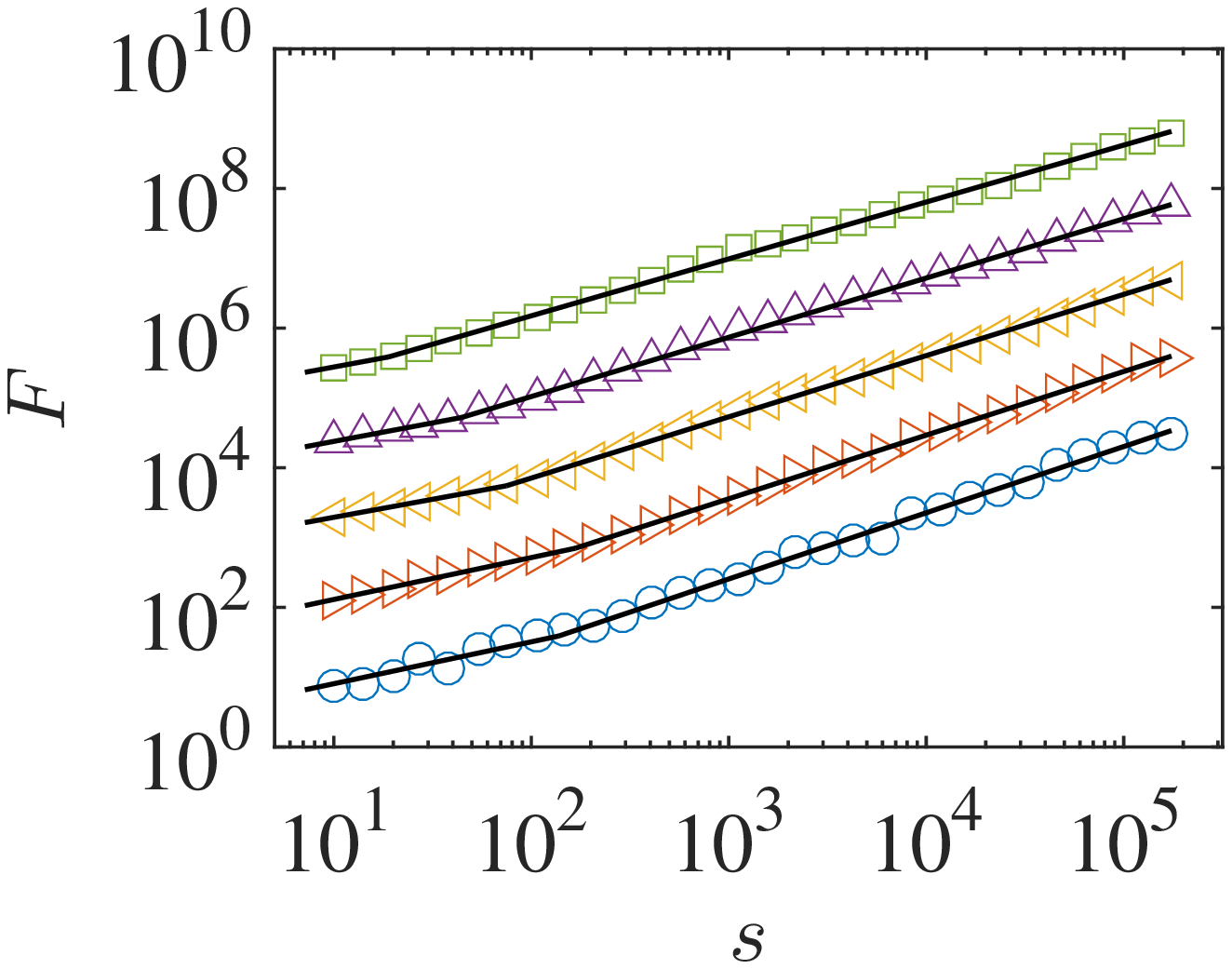}
  \includegraphics[width=4cm]{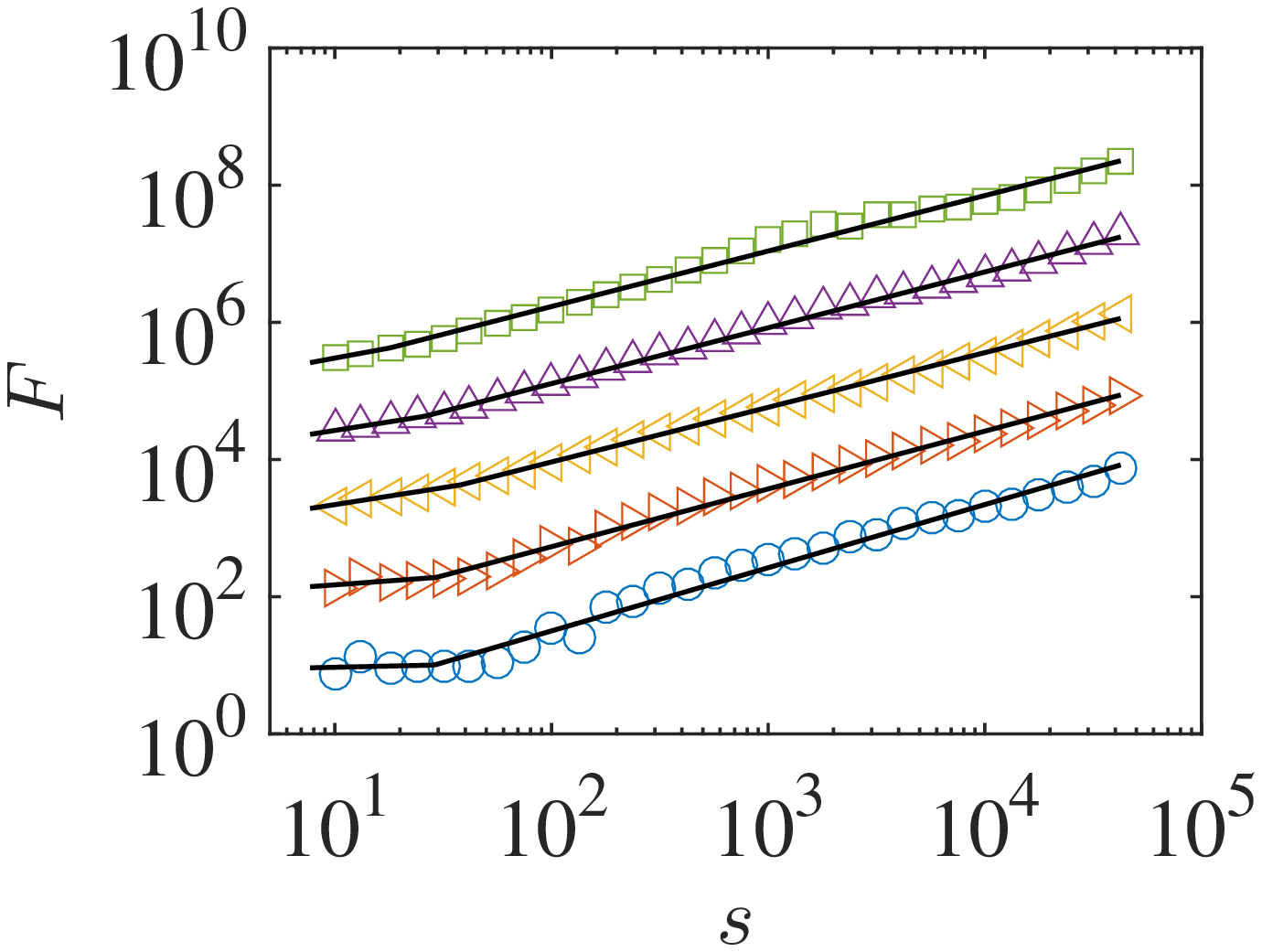}\vskip 0.3 cm
  \includegraphics[width=4cm]{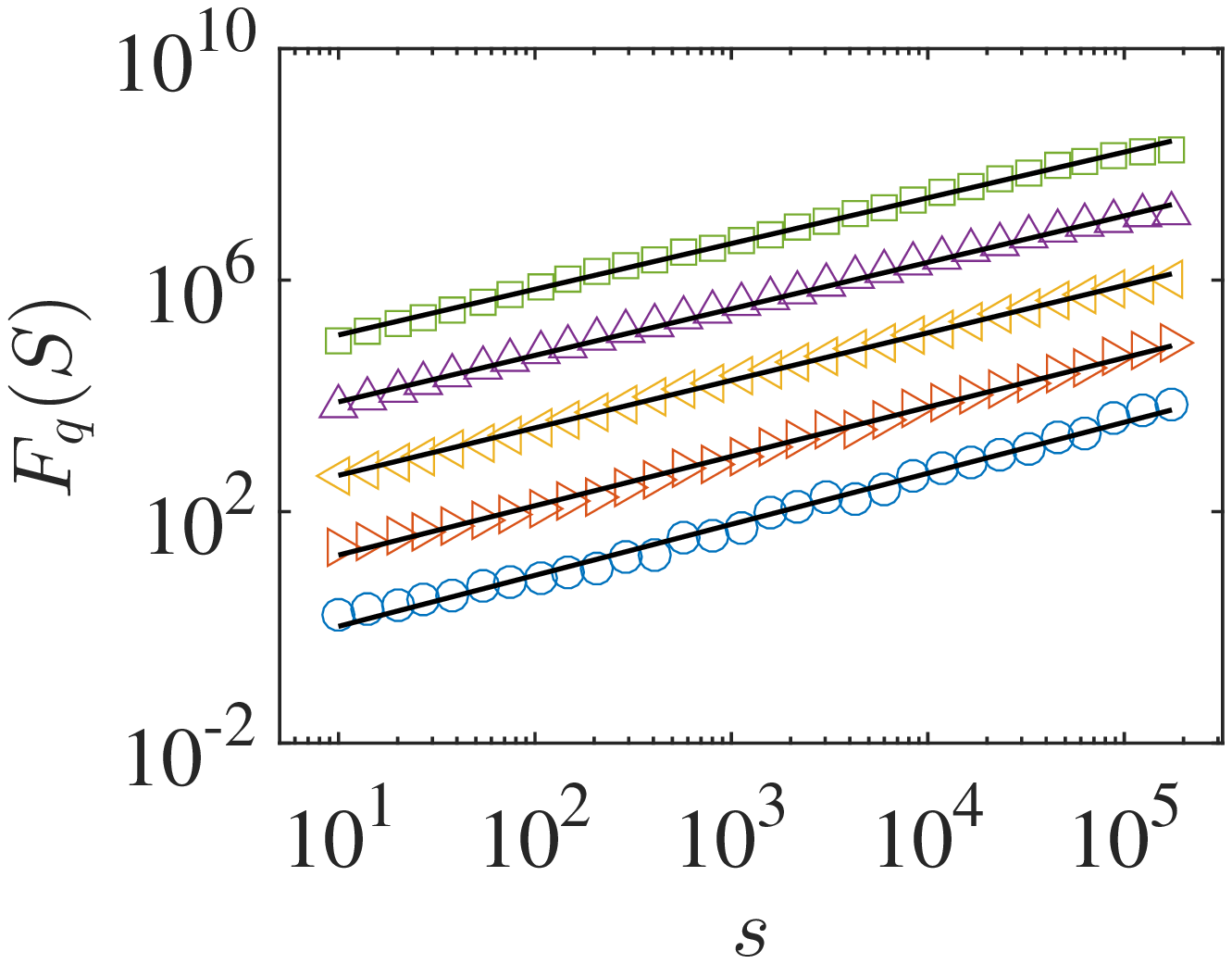}
  \includegraphics[width=4cm]{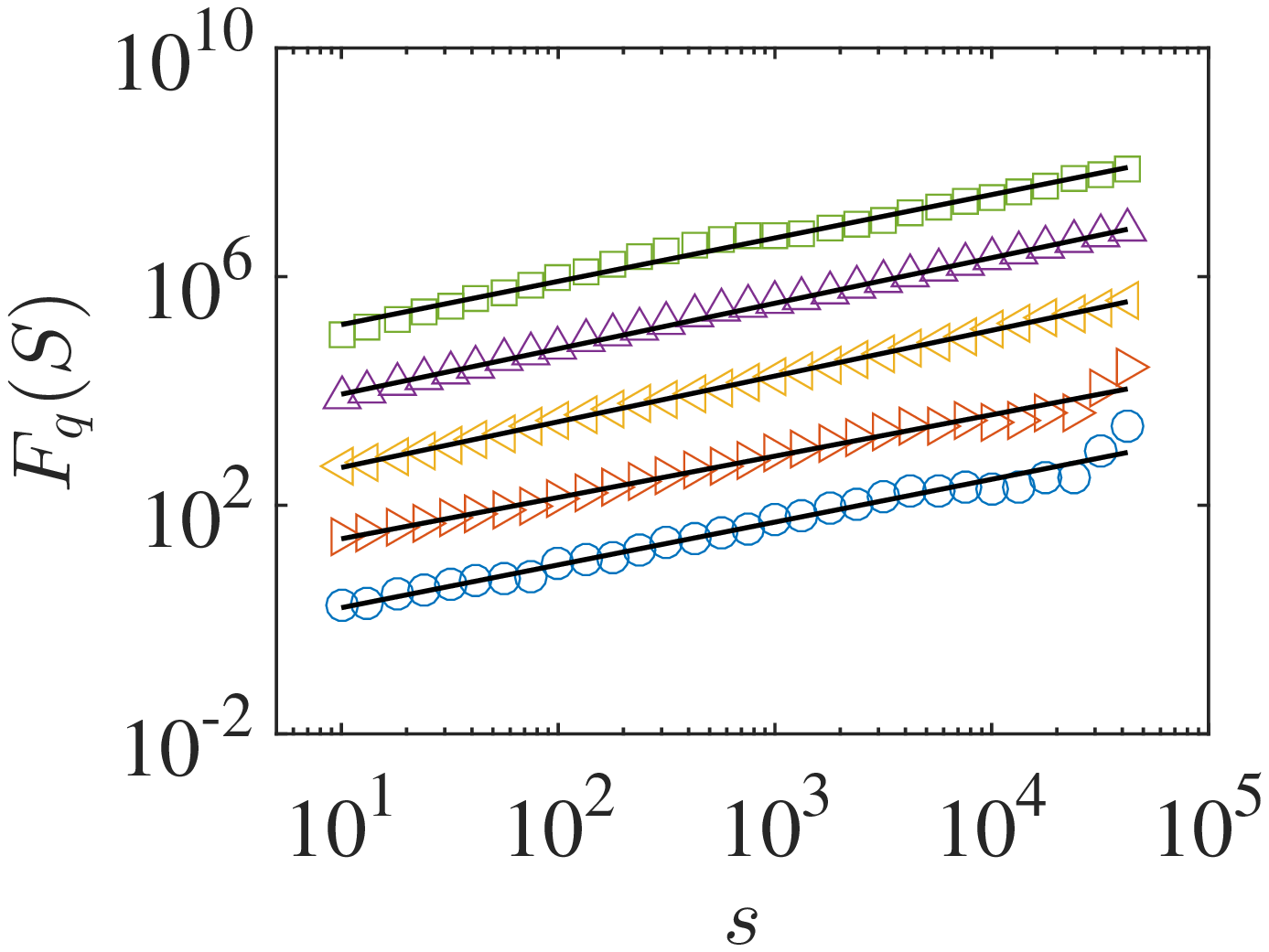}
  \caption{(color online) Power-law dependence of the fluctuation functions $F_q(s)$ with respect to the scale $s$ for $q=-10$, $q=-5$, $q=0$, $q=5$ and $q=10$, using the backward (top row), centered (middle row) and forward (bottom row) MFDMA methods. The straight lines are the best power-law fits to the data. The results have been translated vertically for better visibility. The left column is for stock 000009 and the right column is for stock 000024.}
  \label{Fig:MFDMA:Agg:Fq:s}
\end{figure}

Figure \ref{Fig:MFDMA:Agg:Fq:s} illustrate the dependence of the fluctuation function $F_q(s)$ with respect to the scale $s$ for two stocks 000009 and 000024, using the backward, centered and forward MFDMA methods. For the results from the backward and forward MFDMA methods, we observe nice power-law scalings without clear crossovers. In contrast, the curves from the centered MFDFA exhibit evident crossovers.

\begin{figure}[htb]
  \centering
  \includegraphics[width=4cm]{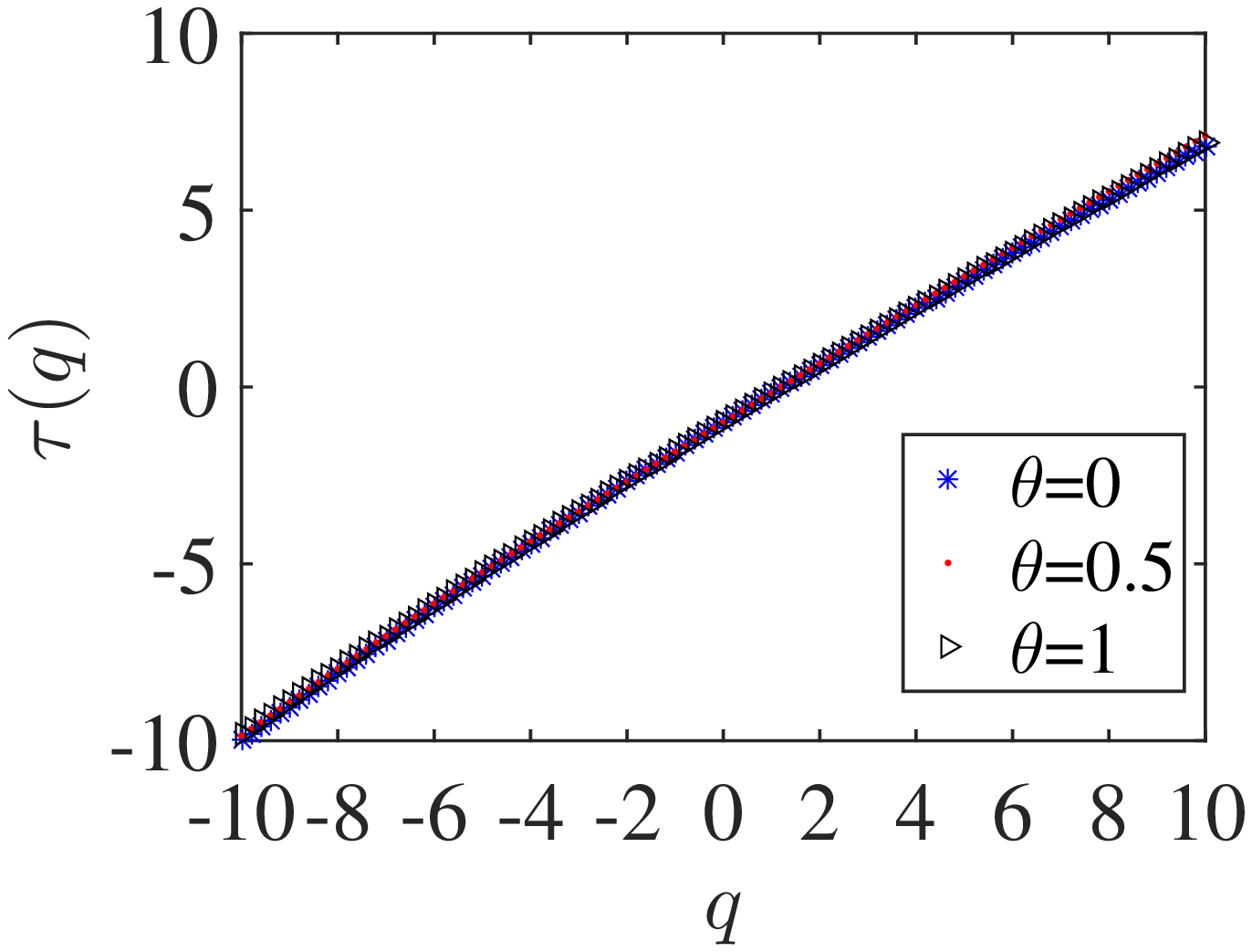}
  \includegraphics[width=4cm]{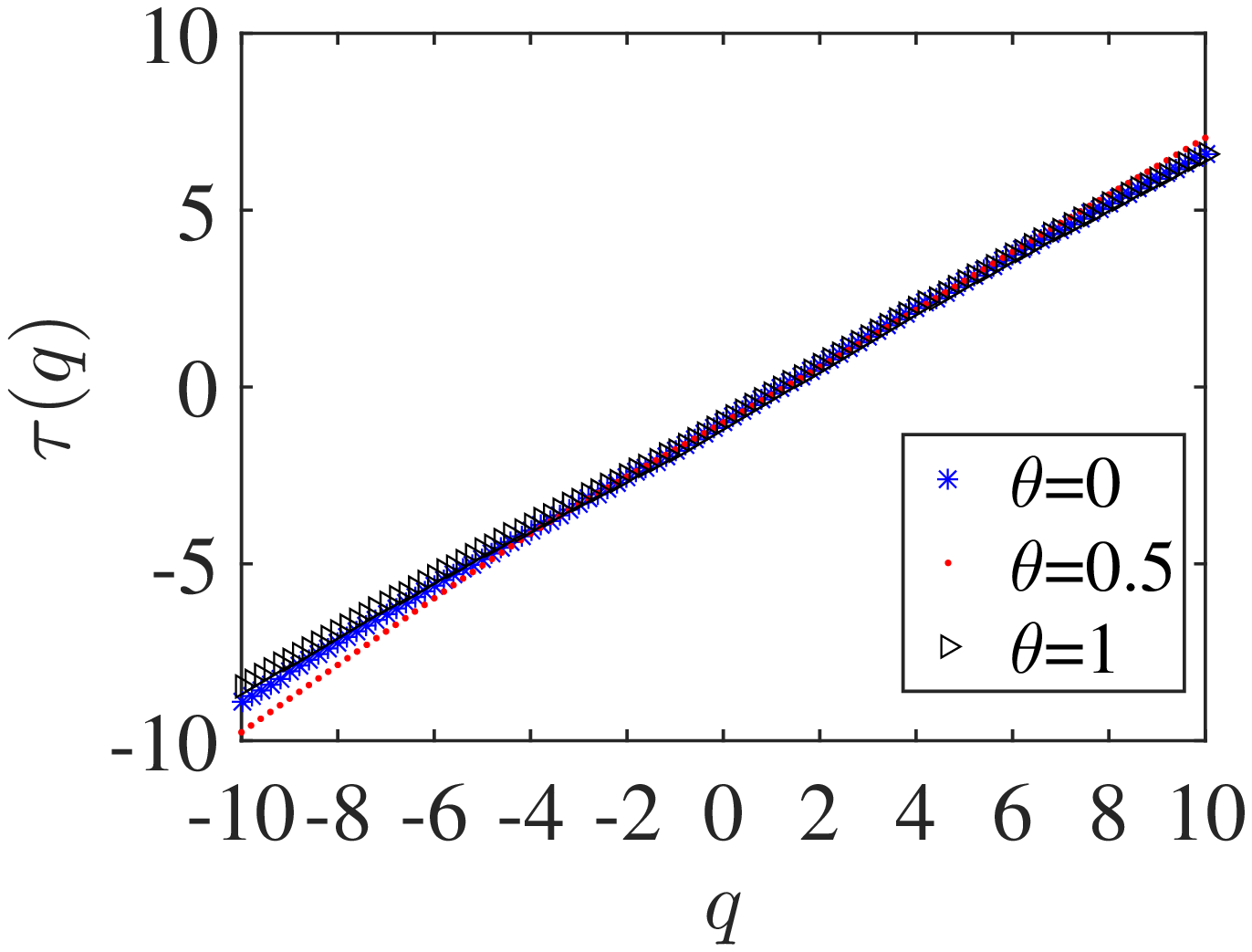}\vskip 0.2 cm
  \includegraphics[width=4cm]{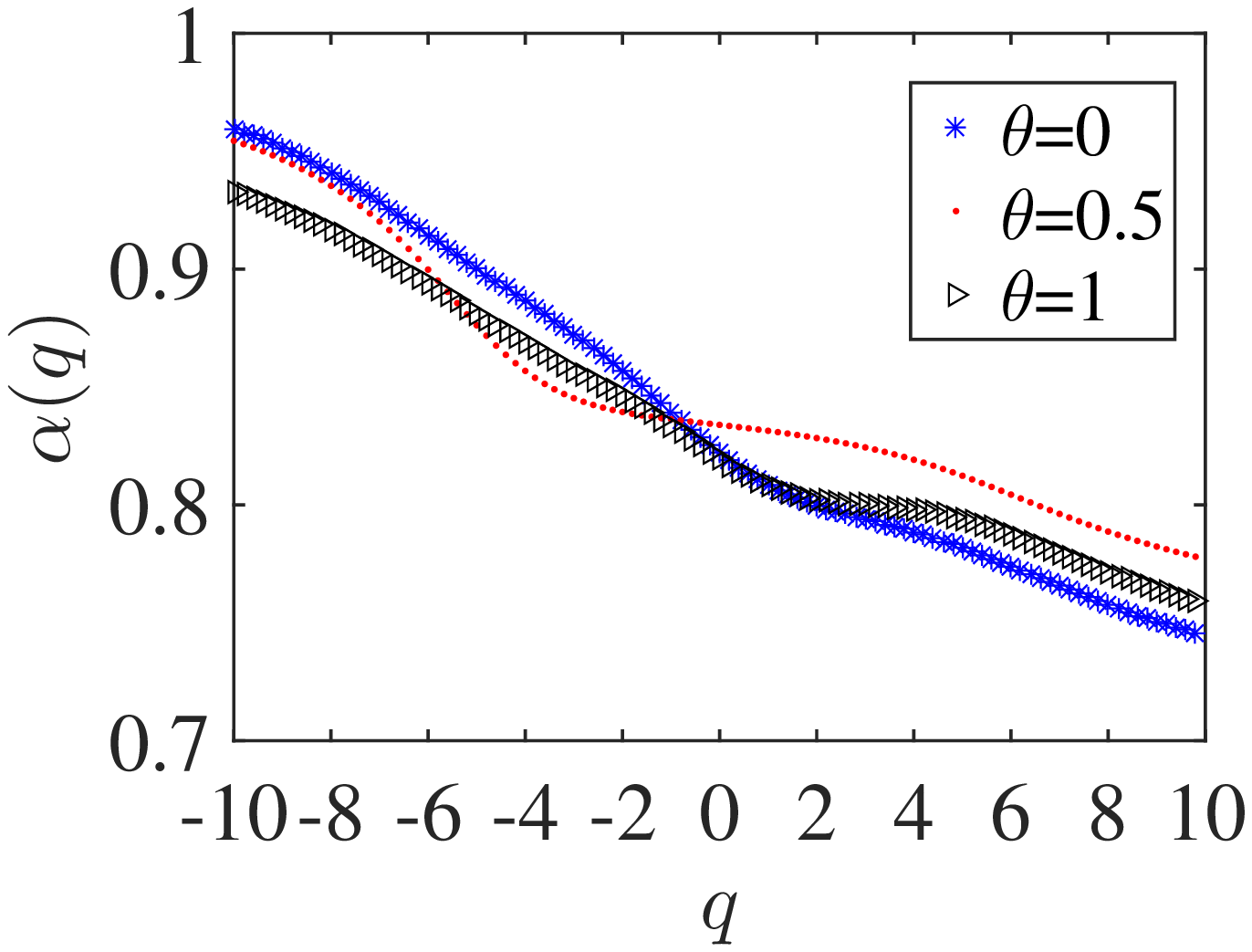}
  \includegraphics[width=4cm]{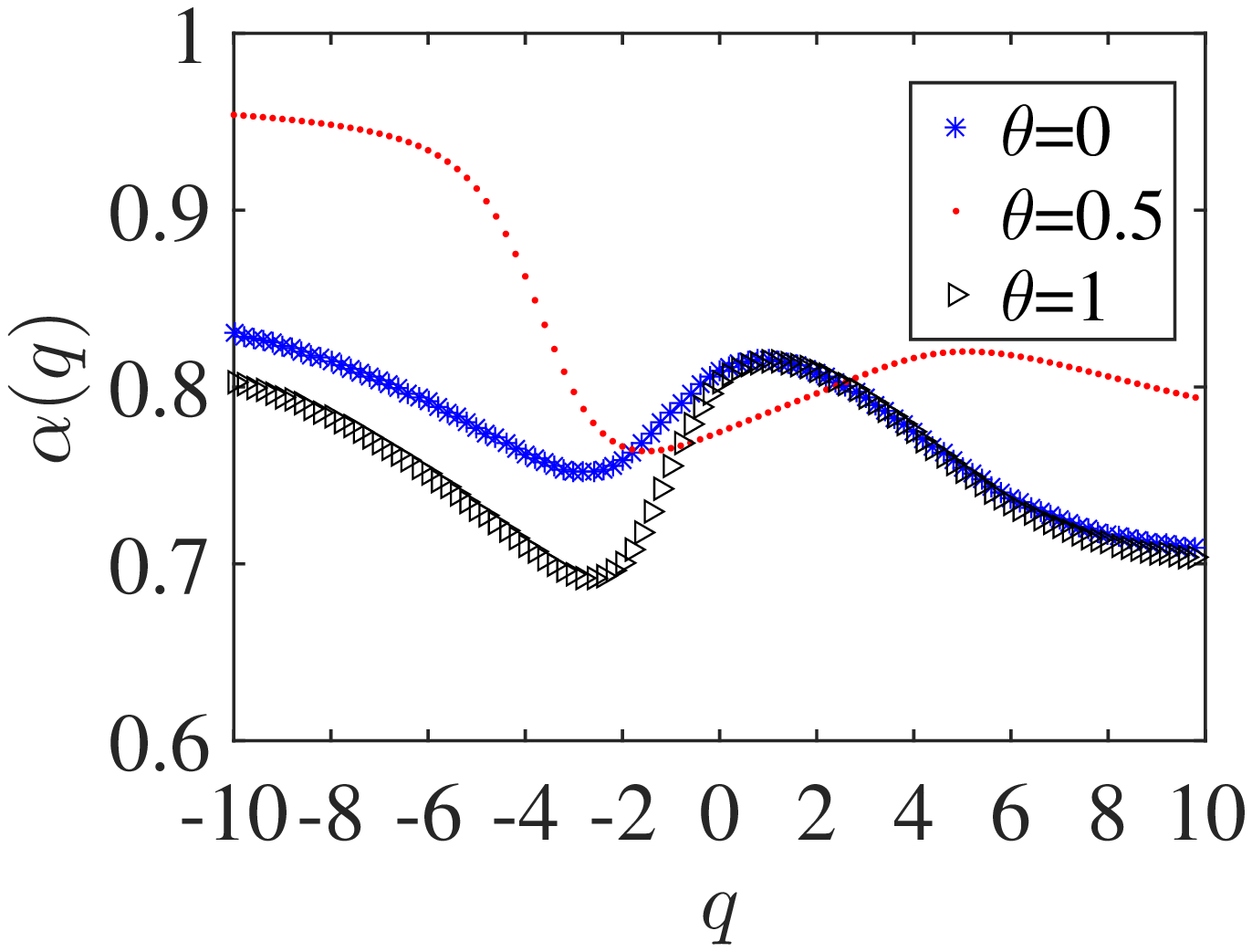}\vskip 0.2 cm
  \includegraphics[width=4cm]{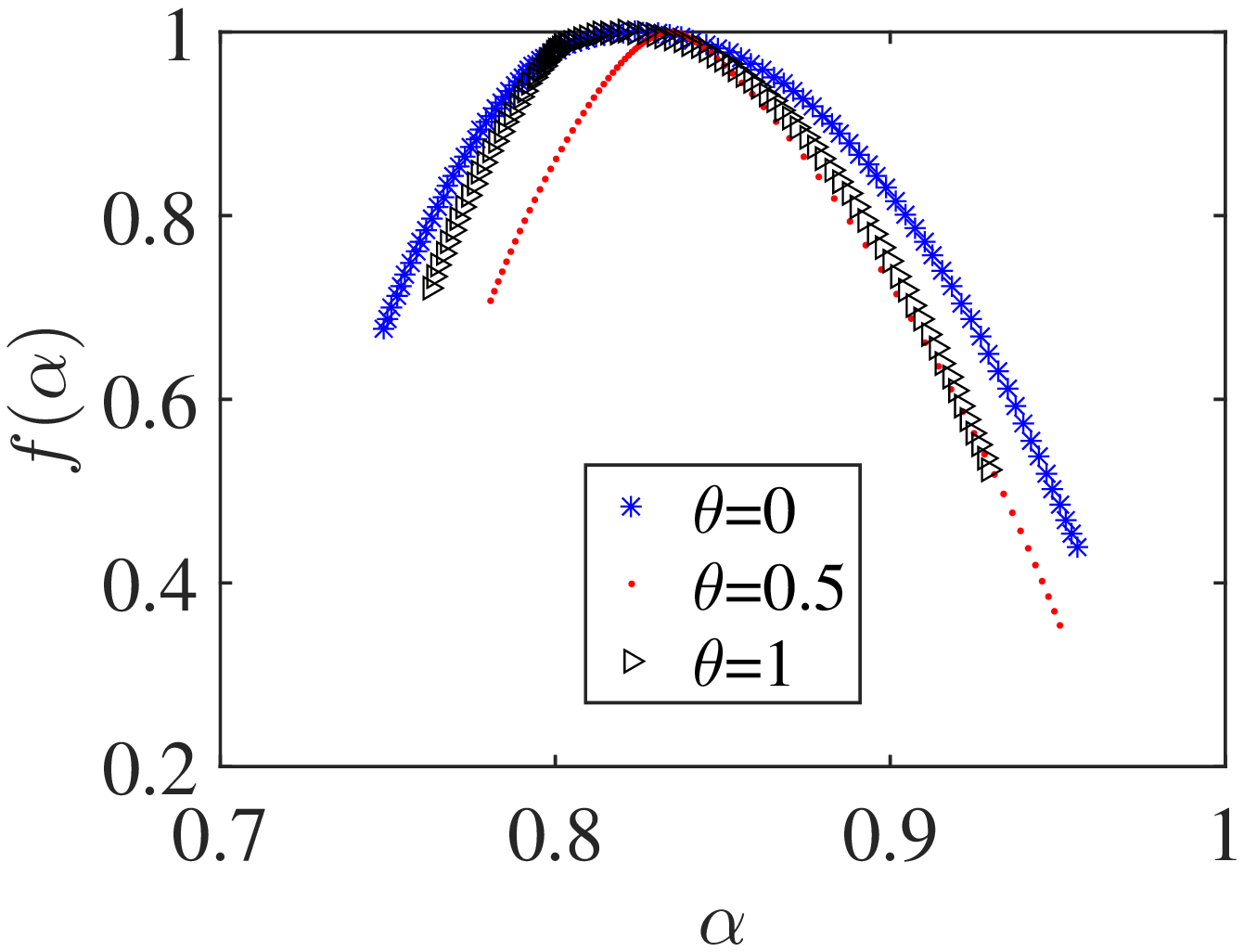}
  \includegraphics[width=4cm]{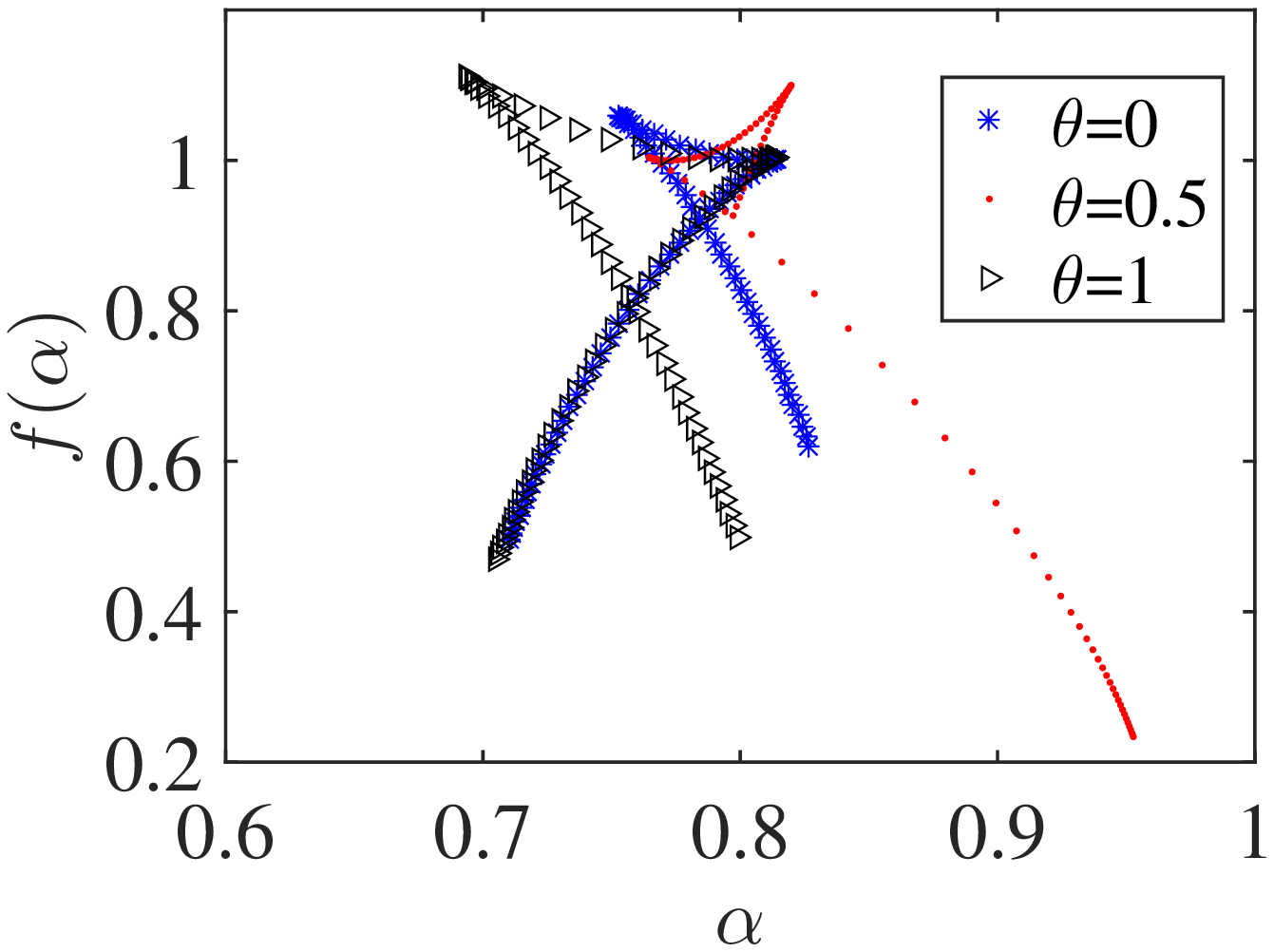}
  \caption{Estimated mass exponent functions $\tau(q)$, singularity strength functions $\alpha(q)$ and singularity spectra $f(\alpha)$ for stock 000009 (left column) and stock 000024 (right column).}
  \label{Fig:MFDMA:Agg:MF}
\end{figure}

According to the standard multifractal formalism, the multifractal scaling exponent $\tau(q)$ can be used to characterize the multifractal nature, which reads
\begin{equation}
  \tau(q)=qh(q)-D_f,
  \label{Eq:tau:hq}
\end{equation}
where $D_f$ is the fractal dimension of the geometric support of the measure \cite{Kantelhardt-Zschiegner-KoscielnyBunde-Havlin-Bunde-Stanley-2002-PA}. We have $D_f=1$ for time series. If the mass exponent function $\tau(q)$ is a concave function of $q$, the measure has multifractal nature. Plots (a) and (b) of Fig.~\ref{Fig:MFDMA:Agg:MF} show the mass exponent functions $\tau(q)$ for the two stocks. It is found that the three curves exhibit discrepancy and do not overlap for each stock. For stock 000009, the $\tau(q)$ functions are seemingly concave. However, for stock 000024, the $\tau(q)$ functions show abnormal curvatures around $q=0$. Hence, the nonlinearity of the $\tau(q)$ function does not ensure that the time series has a multifractal nature. One needs to investigate if the singularity spectrum has a bell-like shape.

The singularity strength function $\alpha(q)$ and the multifractal spectrum $f(\alpha)$ can be obtained numerically through the Legendre transform \cite{Halsey-Jensen-Kadanoff-Procaccia-Shraiman-1986-PRA}
\begin{equation}
    \left\{
    \begin{array}{ll}
        \alpha(q)={\rm{d}}\tau(q)/{\rm{d}}q\\
        f(q)=q{\alpha}-{\tau}(q)
    \end{array}
    \right..
\label{Eq:f:alpha:tau}
\end{equation}
In Fig.~\ref{Fig:MFDMA:Agg:MF}(c) and (d), we illustrate the singularity strength functions $\alpha(q)$. It is obvious that the shape of the $\alpha(q)$ functions is very different from the conventional curves of multifractal measures in which $\alpha(q)$ is a monotonically decreasing function. In Fig.~\ref{Fig:MFDMA:Agg:MF}(e) and (f), we show the singularity spectra $f(\alpha)$. The $f(\alpha)$ curves seem normal for stock 000009 but abnormal for stock 000024. Overall, the results indicate that the order aggressiveness time series possess multifractal nature and some stocks seem to have richer dynamics than others \cite{Czarnecki-Grech-2010-APPA}.

\section{Conclusion}
\label{S1:Conclusion}

In this work, we investigated the linear and nonlinear long-range correlations in the time series of order aggressiveness of 43 Chinese stocks, using the detrending moving average analysis and the multifractal detrending moving average analysis. The DMA analysis identified crossovers in the scaling behaviors of overall fluctuations and the order aggressiveness time series has been found to exhibit linear long-term correlations. We found that, the short-term correlation is independent of the key firm-specific characteristics, while the long-term correlation strength increases with daily turnover rate and decreases with income per share. The stronger long-term correlations and the associated higher turnover rate are both caused by the stronger imitations among traders \cite{Biais-Hillion-Spatt-1995-JF}.

The multifractal detrending moving average analysis shows that the order aggressiveness time series exhibit nice scaling laws of the fluctuation functions and the singularities broadly distributed. It means that these time series possess multifractality. However, the behaviour of some stocks (like stock 000024) are different from others (like stock 000009) in the sense that their singularity spectra are not bell-shaped. It suggests that stock 000024 reveals much more dynamic behaviour with extreme events while stock 000009 is more stable and ``quiet'' \cite{Czarnecki-Grech-2010-APPA}. The asymmetry of the multifractal spectrum towards the right side in Fig.~\ref{Fig:MFDMA:Agg:MF}(a) is observed in many empirical systems \cite{Drozdz-Oswiecimka-2015-PRE}, which might implies different underlying mechanisms. However, the the situation is very complicated because the multifractal spectrum can be right-sided in one market \cite{Oswiecimka-Kwapien-Drozdz-2005-PA} but left-sided in another market \cite{Ruan-Zhou-2011-PA}.

\begin{acknowledgments}
  We acknowledge financial supports from the National Natural Science Foundation of China (71532009, 71501072, 71571121 and 71671066) and the Fundamental Research Funds for the Central Universities (222201718006).
\end{acknowledgments}

\bibliographystyle{naturemag}
\bibliography{E:/Papers/Auxiliary/Bibliography}

\end{document}